\documentclass[12pt]{iopart}
\newcommand{\bsigma}{\mbox{\boldmath $\sigma$}}

\newcommand{\bdeta}{\mbox{\boldmath $\eta$}}

\newcommand{\sgn}{ ~{\rm sgn} }

\newcommand{\be}{\begin{equation}}
\newcommand{\ee}{\end{equation}}
\newcommand{\bd}{\begin{displaymath}}
\newcommand{\ed}{\end{displaymath}}
\newcommand{\bra}{\langle}
\newcommand{\ket}{\rangle}

\newcommand{\here}{\makebox(0,0)}
\usepackage{amssymb}
\usepackage{epsfig}
\usepackage{graphicx}

\begin{document}
\title{Dynamical replica theoretic analysis of CDMA detection dynamics}
\author{J.P.L. Hatchett$^\dag$ and M. Okada$^\S$}
\address{\dag~ Laboratory for mathematical neuroscience, RIKEN Brain
  Science Institute, Hirosawa 2-1, Wako Shi, Saitama 351 0198, Japan}
\address{\S~ Department of Complexity Science and Engineering,
  University of Tokyo, Kashiwanoha 5-1-5, Kashiwa, Chiba, 277-8561,
  Japan, PRESTO JST, Japan}

\begin{abstract}
We investigate the detection dynamics of the Gibbs sampler for
code-division multiple access (CDMA) multiuser detection. Our approach
is based upon dynamical replica theory which allows an analytic
approximation to the dynamics. We use this tool to investigate the
basins of attraction when phase coexistence occurs and examine its efficacy via
comparison with Monte Carlo simulations.
\end{abstract}

\ead{\tt hatchett@brain.riken.jp, okada@k.u-tokyo.ac.jp}

\section{Introduction}
Mobile phone communication is now a key technology across large
swathes of the world. One of the essential technological ingredients
for its success is the ability for multiple users to share a single
channel (i.e. many people can use the same channel to communicate
between their mobile phones and a particular base station). Code-division
multiple access \cite{Simonetal94, Viterbi95, Verdu98} is a protocol that allows this 
multiple access to a single channel through each user modulating
their signal (via so called spreading codes) before transmitting to
the base station. The base station 
receives a mixture of these modulated signals, combined with
aditional channel noise, and the task is then to use knowledge of the
spreading codes and received signal to reconstruct the original
information. 
CDMA has been the subject of several studies in the last few years that have
utilised the relationship between a model of the communication process
and the statistical mechanics of fully connected disordered Ising spin
systems to examine the posterior distribution of the orignal signal using
Bayesian inference \cite{Tanaka01, Tanaka02, Nishimori01,
  Nishimori02}. This gives access to maximum a posteriori (MAP) and maximum
posterior marginal (MPM) decoding. Progress has also been made in terms
of algorithms for decoding using message passing procedures
\cite{Kabashima03, Saad05, Neirotti05}. Recently, both density
evolution and generating functional analysis have been used to
analysze the {\em dynamics} of some detection algorithms for the parallel inference
canceller \cite{Tanaka05, Mimura05}. The first of these techniques
makes the relatively strong approximation of a Gaussian local field,
and ignores the Onsager reaction term in the local field, however it
is known to generally give relatively good results when the detection dynamics
converge \cite{Tanaka05}. In contrast, the generating functional
approach \cite{Dedominicis78, Mimura05} is {\em exact}, but the complexity both
analytically and numerically increases rapidly with the number of time
steps considered and thus it is only practically useful for examining the first
few steps of the dynamics. 
In the current paper we exploit the alternative approach of
dynamical replica theory \cite{Coolen94, Coolen94a, Laughton96}. This
allows us to treat the dynamics of a sequential update detection
algorithm (namely the Gibbs sampler) working in continuous time, with
an analytic approximation scheme that we expect to be superior to
density evolution and with a numerical effort that increases only
linearly in time. It was noted in \cite{Tanaka01} that there exists a
spinodal for both MPM and MAP decoding, past which the decoding
problem has two locally stable solutions, one with good performance
and one with relatively poor performance. This coexistence has
practical implications since local search algorithms starting from an
initial state with a relatively high error rate are closer to the
poor solution (at least in the sense of the error rate). To go beyond
this qualitative argument, however, one really requires dynamical
tools since the basins of attraction are dynamically defined
concepts. The theory which we develop in the current paper allows us
to examine these concepts. We examine the theory for the Gibbs
sampler, as a prototype local search algorithm, not because we believe
that it is necessarily the optimal algorithm for detection and
decoding in CDMA type problems. The dynamical replica approach we
describe here is an approximation, but its justification is given by
comparison with Monte Carlo simulations. Finally we note that the
theory we develop here is also applicable with minor modifications to
the linear Ising perceptron \cite{Neirotti05}

\section{Model definitions and order parameter evolution equations}
We consider the demodulation problem for the $N$-user direct-sequence
binary phase-shift-keying (DS/BPSK) CDMA system, with the simplifying
assumptions that the channel noise is additive white Gaussian, chip
and symbol timing are perfectly synchronized across users and the
output power of the users is perfectly equalized by power
control. For details of the equilibrium statistical mechanical
analysis of this model please see \cite{Tanaka01, Tanaka02,
  Nishimori01} .

We consider $N$ users sending information bits $\sigma_i^0 \in
\{-1,1\}\ \forall i=1,\ldots,N$. Each user $i$ has a binary spreading code
$\{\eta_i^t:t=1,\ldots p\},\ \eta_i^t \in \{-1,1\}$ so that in symbol
interval $t$, user $i$ transmits $\eta_i^t \sigma_i^0$. We model the
speading code sequences to be independent quenched random variables with Prob$[\eta_i^t =
  \pm 1] = \frac12$ and take the zero mean additive white Gaussian noise $\{\nu^t:t =
1,\ldots,p\}$ to have variance $N/\beta_s$. Thus, at each chip time step
$t \in \{1,\ldots, p\}$ the received signal at the base station $y^t$ is given by 
\begin{eqnarray}
y^t = \sum_{i=1}^N \eta_i^t \sigma^0_i + \nu^t
\end{eqnarray}
Bayesian inference shows that posterior distribution of the original
signal, given the noisy signal is given by a Gibbs-Boltzmann
distribution with temperature $\beta$ and Hamiltonian
\begin{eqnarray}
\fl H(\bsigma) =  \frac12\sum_{ij} J_{ij} \sigma_i \sigma_j -
\sum_{i=1}^N f_i \sigma^0_i \qquad J_{ij} = \frac{1}{N} \sum_t \eta_i^t
\eta_j^t \qquad f_i = \frac1N \sum_{t=1}^p \eta_i^t y^t
\end{eqnarray}
where the signal $\{y^t\}$ and the spreading codes $\{\eta_i^t\}$
constitute quenched disorder. The temperature $\beta$ is a free
control parameter,  MAP decoding corresponds to
the limit $\beta \to \infty$ while MPM decoding corresponds to
$\beta = \beta_s$ (the Nishimori temperature
\cite{Nishimori01}), although in general $\beta_s$ may not be known.

We examine the detection problem by using a spin system with
Glauber dynamics to model the posterior distribution. To study the
dynamical evolution of this distribution analytically, we use the techniques of dynamical replica theory
\cite{Coolen94, Coolen94a,  Laughton96}, at the level of a three parameter
approximation. With Glauber dynamics the time evolution of the
microscopic state probability distribution $p_t(\bsigma)$ is given by
the master equation
\begin{eqnarray}
\frac{\rmd}{\rmd t} p_t(\bsigma) = \sum_{k=1}^N[p_t(F_k\bsigma)
  w_k(F_k\bsigma) - p_t(\bsigma) w_k(\bsigma)] \label{eq:Master}
\end{eqnarray}
with the spin-flip operator $F_k \Phi(\bsigma) \equiv
\Phi(\sigma_1,\ldots,-\sigma_k,\ldots,\sigma_N) $ and transition
rates $w_k(\bsigma)$ given in terms of the local alignment fields
$h_k(\bsigma)$ as
\begin{eqnarray}
w_k(\bsigma) = \frac12[1 -\sigma_k \tanh(\beta h_k(\bsigma))]\qquad
h_k(\bsigma) = f_k - \sum_{j \neq k} J_{kj} \sigma_j
\end{eqnarray}
Conventional demodulation \cite{Verdu98} corresponds to taking $\hat{\sigma}^0_k =
\mbox{sgn}(f_k)$ where $\hat{\bsigma}^0$ is our estimator for $\bsigma$, the
true signal. To improve upon this we take into account
correlations induced by the spreading code. The dynamics
(\ref{eq:Master}) lead asymptotically to the required posterior
distribution in the high $t$ limit (i.e. the Bayesian posterior
distribution). The primary performance 
measure for any demodulator is given by the overlap $M$ between the signal
$\bsigma^0$ and the estimate of the signal $\bsigma$, defined by
\begin{eqnarray}
M(\bsigma) = \frac{1}{N} \sum_i \sigma_i \sigma_i^0
\end{eqnarray}
We also use as macroscopic order parameters the internal energy
\begin{eqnarray}
E(\bsigma) = \frac{1}{2N} \sum_{ij} \sigma_i J_{ij} \sigma_j = \frac{\alpha}{2}
+ \frac{1}{2N} \sum_{i \neq j} \sigma_i J_{ij} \sigma_j
\end{eqnarray}
(note the similarity to the order parameter
$r(\bsigma)$ from \cite{Coolen94a}) and the contribution due to the
external fields 
\begin{eqnarray}
F(\bsigma) = \frac{1}{N} \sum_i f_i \sigma_i
\end{eqnarray}
The first order parameter is our performance measure while the latter two
give the energy from the Hamiltonian. We have chosen to split the
energetic term into two pieces since we will find that under our
assumptions both $E$ and $F$ will evolve according to odes containing
a relaxation term and a complicated force term. Since, $E$ is
quadratic in the spins, and $F$ is linear in the spins, if we took
$E-F$ (i.e. the energy of the system) to be a single order paramter,
then its evolution would follow 
from the difference of two complicated force terms. We found that the
degree of analytic complexity was the same for either choice. However,
by splitting the energy into three terms our approximation to
$p_t(\bsigma)$ has an extra degree of freedom and thus the
approximation is better. Hence, at no extra analytic or numerical cost
(compared to the standard two order parameter theory) we obtain a
better approximation to the dynamics. 

Following \cite{Coolen94, Coolen94a, Laughton96} we 
may derive a Kramers-Moyal expansion for the probability density
$\mathcal{P}_t(M,E,F) = \sum_{\bsigma} p_t(\bsigma) \delta[M - M(\bsigma)]
\delta[E - E(\bsigma)] \delta[F - F(\bsigma)]$,
\begin{eqnarray}
\fl \frac{\rmd}{\rmd t} \mathcal{P}_t(M,E,F) = -\frac{\partial}{\partial M}
\left\{ \mathcal{P}_t(M,E,F) \left[ \left\langle \frac{1}{N} \sum_i
  \sigma_i^0 \tanh[\beta h_i(\bsigma)] \right\rangle_{M,E,F;t} - M
  \right] \right\} \nonumber\\
-\frac{\partial}{\partial E} \left\{ \mathcal{P}_t(M,E,F) \left[
  \left\langle\frac{1}{N} \sum_{i} 
  h_i^{loc}(\bsigma) 
  \tanh[\beta h_i(\bsigma)] 
  \right\rangle_{M,E,F;t} +\alpha- 2E  \right]  \right\} 
\nonumber\\
-\frac{\partial}{\partial F} \left\{\mathcal{P}_t(M,E,F) \left[
  \left\langle \frac{1}{N} \sum_{i} 
f_i(\bsigma) 
  \tanh[\beta h_i(\bsigma)]
  \right\rangle_{M,E,F;t} - F   \right] \right\} 
\nonumber\\
+ \mathcal{O}(\frac1N)
\end{eqnarray}
where we define $h_i^{loc}(\bsigma) = \sum_{j \neq i} J_{ij}
\sigma_j$. In the thermodynamic limit, on finite timescales, only the Liouville term survives  
in this equation, so that the order parameter triple $(M,E,F)$ evolves
deterministically according to
\begin{eqnarray}
\frac{\rmd}{\rmd t} M = -M + \left\langle \frac{1}{N} \sum_i
\sigma_i^0 \tanh[\beta h_i(\bsigma)] \right\rangle_{M,E,F;t}\label{eq:M_ode}\\
\frac{\rmd}{\rmd t} E = -2E + \alpha + \left\langle\frac{1}{N} \sum_{i}
h_i^{loc}(\bsigma) \tanh[\beta h_i(\bsigma)] \right\rangle_{M,E,F;t} \label{eq:E_ode} \\
\frac{\rmd}{\rmd t} F = -F + \left\langle\frac{1}{N} \sum_{i}
f_i \tanh[\beta h_i(\bsigma)] \right\rangle_{M,E,F;t} \label{eq:F_ode}
\end{eqnarray}
where
\begin{eqnarray}
\fl \left\langle f(\bsigma) \right \rangle_{M,E,F;t} = \frac{\sum_{\bsigma}
  p_t(\bsigma) \delta[M - M(\bsigma)] \delta[E - E(\bsigma)] \delta[F
  - F(\bsigma)]
  f(\bsigma)}{ \sum_{\bsigma}
  p_t(\bsigma) \delta[M - M(\bsigma)] \delta[E - E(\bsigma)]\delta[F
  - F(\bsigma)]}\label{eq:unpleasant}
\end{eqnarray}
This flow equation is still exact in the thermodynamic limit. However,
to move to a practical representation, i.e. one that does not depend
on the microstate probability distribution $p_t(\bsigma)$, we make the assumptions
underlying dynamical replica theory \cite{Coolen94, Laughton96}: that
the observables are self-averaging with respect to the realisation of
the disorder and initial conditions  and that we may appoximate the
microscopic probability distribution by the maximum entropy
distribution given the values of our observables. Thus, with this
approximation the microstate probability drops out of our equations
and we may then use the replica technique to remove the unpleasant
fraction in (\ref{eq:unpleasant}), via
\begin{eqnarray}
\frac{\sum_{\bsigma} \Phi(\bsigma) W(\bsigma)}{ \sum_{\bsigma}
  W(\bsigma)} = \lim_{n \to 0} \sum_{\bsigma^1,\ldots,\bsigma^n}
  \Phi(\bsigma^1) \prod_{\alpha = 1}^n W(\bsigma^\alpha)
\end{eqnarray}
which we proceed to do in the following section.

\section{Replica calculation of the flow}
Using site equivalence under the disorder average,  the objects we
need to calculate can be expressed as 
\begin{eqnarray}
D(h) = \overline{\bra \sigma^0_1 \delta[h - h_1(\bsigma)]\ket_{M,E,F} }\\
D(f,h) =  \overline{ \bra \delta[f - f_1] \delta[h -
    h_1^{loc}(\bsigma)]\ket_{M,E,F} } \label{eq:dfh}
\end{eqnarray}
From the definition $h_1(\bsigma) \equiv f_1 -
h_1^{loc}(\bsigma)$ it is trivial to obtain the first distribution from
the second, so we focus on the calculation of the second term in the
current section.
We introduce Fourier representations for the delta functions over $(M,E,F)$
constraining the distribution in (\ref{eq:dfh}) giving conjugate
parameters $\{\hat{M}^\alpha, \hat{E}^\alpha, \hat{F}^\alpha\}$. To
average over the disordered signal $y^t$, we generate its correct
measure by using the partition function with the true value of the
noise $\beta_s$ \cite{Nishimori01}. We also write the delta functions over
the fields in Fourer representation
\begin{eqnarray}
\fl \delta[f -\frac1N \sum_t y_t \eta_1^t] \delta[h - h_1(\bsigma^1)]
  = \int \frac{\rmd \hat{h} \rmd \hat{f}}{(2\pi)^2}\nonumber\\
  \exp\left\{\rmi 
  h\hat {h} +\rmi f \hat{f} - \frac{\rmi\hat{f}}{N} \sum_{t=1}^p y^t \eta_1^t -
  \frac{\rmi\hat{h}}{N} \sum_{j>1} \sum_{t= 1}^p \eta_1^t \eta_j^t \sigma_j^1
  \right\}
\end{eqnarray}
To get the correct scaling, we change variables $r^t = y^t/\sqrt{N}$
and then introduce
\begin{eqnarray}
1 \sim \prod_{t=1}^p \prod_{\alpha = 0}^n \int \rmd v^{t\alpha} \rmd
\hat{v}^{t\alpha} \exp\left\{\rmi v^{t\alpha}\hat{v}^{t\alpha} -
  \frac{\rmi}{\sqrt{N}} \hat{v}^{t\alpha} \sum_{i > 1} \eta_i^t
  \sigma_i^\alpha\right\} 
\end{eqnarray}
We then find
\begin{eqnarray}
\fl D(f, h) \sim  \lim_{n \to 0} \int \frac{\rmd \hat{h} \rmd \hat{f}}{(2\pi)^2}
\prod_\alpha [\rmd 
\hat{M}^\alpha \rmd \hat{E}^\alpha \rmd \hat{F}^\alpha] \prod_{t=1}^p \rmd r^t
\prod_{t\alpha} [\rmd v^{t\alpha} \rmd \hat{v}^{t\alpha}] 
\rme^{\rmi h\hat{h} + \rmi \hat{f}f+\rmi N \sum_\alpha(M \hat{M}^\alpha + E
\hat{E}^\alpha) }\nonumber \\
\fl\rme^{\rmi N \sum_\alpha  F \hat{F}^\alpha + \rmi \sum_{\alpha t}
v^{t\alpha}\hat{v}^{t\alpha} -\frac{\beta_s}{2} \sum_t (r^t -
v^{t0})^2 -\frac{\alpha 
\beta_s}{2} -\frac{\rmi}{2} \sum_\alpha \hat{E}^\alpha \sum_t 
v_{t\alpha}^2 -\frac{\rmi\alpha}{2} \sum_\alpha \hat{E}^\alpha
  - \rmi \sum_{\alpha t} \hat{F}^\alpha r^t v^{t\alpha} } \nonumber\\
\fl \sum_{\bsigma^0,\ldots,\bsigma^n} 
\rme^{-\rmi \sum_{\alpha  > 0} \hat{M}^\alpha \sum_{i} \sigma^\alpha_i
  \sigma_i^0} \sum_{\bdeta} \rme^{- \frac{\rmi
}{\sqrt{N}}\sum_{t=1}^p\eta_1^t ( \hat{f}r^t + \hat{h} 
 v^{t1})-
\frac{\rmi}{\sqrt{N}} \sum_{t\alpha}
\hat{v}^{t\alpha} \sum_{i>1} \eta_i^t \sigma_i^\alpha}
\nonumber\\
\fl \rme^{
-\frac{\rmi}{\sqrt{N}} \sum_{\alpha t} \hat{F}^\alpha r^t \eta_1^t
\sigma_1^\alpha
+\frac{\beta_s}{\sqrt{N}} \sum_t \eta_1^t \sigma_1^0 (r^t - v^{t0})-
\frac{\rmi}{\sqrt{N}} \sum_{t\alpha} \hat{E}^\alpha \eta_1^t
\sigma_1^\alpha v^{t\alpha}} \nonumber 
\end{eqnarray}
By $\sim$ we mean that this is correct up to irrelevant normalisation constants
which can be recovered later by dividing through
by $\overline{\bra 1 \ket_{M,E,F}}$ or requring $\int \rmd f \rmd h
D(f,h) = 1$.
Performing the trace over $\bdeta$ in the last line, and then
expanding the resultant formulas, gives a contribution to leading
order in $N$ of
\begin{eqnarray}
\fl \exp\left\{\frac{1}{2N}\sum_{t} [\beta_s \sigma_1^0 (r^t -
    v^{t0}) -\rmi \sum_\alpha \sigma_1^\alpha (
  \hat{E}^\alpha  v^{t\alpha} + \hat{F}^\alpha 
    r^t )
 - \rmi \hat{f} r^{t} -    \rmi\hat{h} v^{t1} ]^2 \right.\nonumber\\
\left. - \frac{1}{2N}
  \sum_{t,i>1,\alpha,\beta} \hat{v}^{t\alpha} \sigma_i^\alpha
  \hat{v}^{t\beta} \sigma_i^\beta\right\}
\end{eqnarray}
We then introduce the Edwards-Anderson order parameter:
\begin{eqnarray}
1= \int \prod_{\rho < \tau} \rmd r_{\rho \tau} \rmd q_{\rho \tau}
\rme^{\rmi \sum_{\rho \tau} r_{\rho \tau} \sum_{i > 1} \sigma_i^\rho
  \sigma_i^\tau - \rmi N \sum_{\rho \tau} r_{\rho \tau}q_{\rho \tau}}
\end{eqnarray}
leading to
\begin{eqnarray}
\fl D(f,h) \sim \lim_{n \to 0} \int \frac{\rmd \hat{h} \rmd \hat{f}}{(2\pi)^2}
\prod_\alpha [\rmd  
\hat{M}^\alpha \rmd \hat{E}^\alpha \rmd \hat{F}^\alpha]\prod_{\rho <
  \tau} [\rmd r_{\rho 
\tau} \rmd q_{\rho \tau}] 
\rme^{\rmi h\hat{h} + \rmi f \hat{f}}\label{eq:dfh2}\\ 
\fl \rme^{ \rmi N \sum_\alpha(M \hat{M}^\alpha + E
\hat{E}^\alpha + F \hat{F}^\alpha)- \rmi N \sum_{\rho \tau} r_{\rho
\tau}q_{\rho \tau} + (N-1) \log \sum_{\bsigma} \rme^{\rmi \sum_{\alpha \beta}
r_{\alpha \beta} \sigma^\alpha \sigma^\beta} -\rmi N \sum_{\alpha
> 0} \hat{M}^\alpha q_{0\alpha}} \sum_{\bsigma_1} \rme^{-\rmi
  \sum_{\alpha} \hat{M}^\alpha \sigma_1^0 \sigma_1^\alpha} \nonumber\\
\fl \prod_{t}\left\{\int \rmd r \rmd \mathbf{v}
G(r,\mathbf{v})\exp\left[\frac{1}{2N} [\beta_s
\sigma_1^0 (r - v^0) - \rmi \sum_\alpha \sigma_1^\alpha (\hat{E}^\alpha 
v^\alpha + \hat{F}^\alpha r) - \rmi \hat{f} r - 
\rmi\hat{h} v^{1}]^2\right] \right\}\nonumber
\end{eqnarray}
Where
\begin{eqnarray}
G(r,\mathbf{v}) = \rme^{-\frac{1}{2}\mathbf{v}\mathbf{q}^{-1}
  \mathbf{v} - \frac{\beta_s}{2} (r-  
  v^0)^2 - \frac{\rmi}{2} \sum_\alpha \hat{E}^\alpha  v_\alpha^2 -
\rmi \sum_\alpha \hat{F}^ \alpha v^\alpha r}
\end{eqnarray}
It is instructive to view (\ref{eq:dfh2}) as a measure over $(f,h)$,
with free parameters which are subsequently averaged over a saddle
point measure. Thus, the free parameters within the order one measure take their
saddle point values. The saddle point surface itself is given by
\begin{eqnarray}
\fl \Psi = \rmi \sum_{\alpha} (M \hat{M}^\alpha + E \hat{E}^\alpha + F
\hat{F}^\alpha) + \log
\sum_{\bsigma} \rme^{\rmi \sum_{\alpha \beta} r_{\alpha \beta}
  \sigma^\alpha \sigma^\beta} - \rmi \sum_{\alpha \beta} r_{\alpha
  \beta} q_{\alpha \beta} -\rmi \sum_{\alpha > 0} \hat{M}^\alpha
q_{0\alpha}\nonumber \\
+ \alpha \log \int \rmd r \rmd \mathbf{v} 
\rme^{-\frac{1}{2}\mathbf{v}\mathbf{q}^{-1}
  \mathbf{v}  - \frac{\beta_s}{2} (r-
  v^0)^2 - \frac{\rmi}{2} \sum_\alpha \hat{E}^\alpha  v_\alpha^2 -
\rmi \sum_\alpha \hat{F}^\alpha v^\alpha r }
\end{eqnarray}

\section{Replica symmetric saddle points}

We make the replica symmetric assumptions, $r_{0\tau} = r,
r_{\rho\tau} = R$, $q_{0\alpha} = m$ and $q_{\alpha \beta} = q$ and
$(\hat{M}^\alpha, \hat{E}^\alpha,\hat{F}^\alpha) = (\hat{M},\hat{E},\hat{F})\
\forall \alpha$. Then, similarly to the original equilibrium
calculation \cite{Tanaka01, Tanaka02, Nishimori01}, we have a saddle point surface  
\begin{eqnarray}
\fl \Psi_{RS} =  \lim_{n \to 0} \frac{1}{n} \Psi
=  \int Dz \log 2 \cosh(\sqrt{R} z + r) - rm - \frac{1}{2}R(1-q) +
\hat{M}(M-m)+  E\hat{E}\nonumber\\ \fl   + F\hat{F} 
+ \alpha \log \int \rmd r \rmd \mathbf{v}
\exp\left\{-\frac12 \mathbf{v} \mathbf{q}^{-1} \mathbf{v} -
\frac{\beta_s}{2} (r- 
  v^0)^2 - \frac{\hat{E}}{2} \sum_\alpha  v_\alpha^2 -
\hat{F} \sum_\alpha  v^\alpha r\right\}
\end{eqnarray}
We wish evaluate the integrals $I$ in the second line of the above so
that we can take the limit $n\to 0$. It is convenient to change variables to 
\begin{eqnarray}
v_0 = u\sqrt{1 - \frac{m^2}{q}} - \frac{tm}{\sqrt{q}}\\
v_\alpha = z_\alpha \sqrt{1-q} - t \sqrt{q}
\end{eqnarray}
where $u,t,\{z_\alpha\}$ are zero mean, uncorrelated, unit variance
Gaussian random variables. Then using the shorthand $Dx =
(2\pi)^{-\frac12}\rme^{-\frac12x^2}$ we have 
\begin{eqnarray}
\fl I = \int Dr Dt Du \exp\left[-\frac{\beta_s}{2} \left(u\sqrt{1 -
    \frac{m^2}{q}} - \frac{tm}{\sqrt{q}} -r \right)^2 \right]\\
\fl \qquad \times \left\{\int Dz \exp\left[-\frac{\hat{E}}{2}
    (z \sqrt{1-q} - t \sqrt{q})^2  - \hat{F} r(z
    \sqrt{1-q} - t \sqrt{q}) \right]\right\}^n\nonumber
\end{eqnarray}
Although cumbersome, these integrals are straightforward and give
\begin{eqnarray}
I = [1 + \beta_s(1 - \frac{m^2}{q})]^{-\frac12}[1 +
  \hat{E}(1-q)]^{-\frac{n} {2}}\frac{1}{\sqrt{ba - c^2}}
\end{eqnarray}
where
\begin{eqnarray}
a = \frac{\beta_s}{1 + \beta_s(1 - \frac{m^2}{q})} -
\frac{n\hat{F}^2(1-q)}{1 + \hat{E}(1-q)}\\
b = 1 + \frac{\frac{\beta_s m^2}{q}}{1+ \beta_s(1 - \frac{m^2}{q})} +
\frac{n\hat{E} q}{1 + \hat{E}(1-q)}\\
c = \frac{\beta_s\frac{m}{\sqrt{q}}}{1 + \beta_s(1 - \frac{m^2}{q})}
 - \frac{n\hat{F}\sqrt{q}}{1 + \hat{E}(1-q)}
\end{eqnarray}
Giving the replica symmetric saddle point surface
\begin{eqnarray}
\fl \Psi_{RS} = \int Dz \log 2 \cosh(\sqrt{R} z + r) - rm - \frac{1}{2}R(1-q) +
\hat{M}(M-m)+  E\hat{E} + F\hat{F} \nonumber\\
-\frac{\alpha}{2} \left\{\log[1 + \hat{E}(1-q)] +
\frac{\beta_s(\hat{E} q + 2m\hat{F} - \hat{F}^2(1-q)) -
  \hat{F}^2(1-q)}{\beta_s[1 + \hat{E}(1-q)]}\right\}
\end{eqnarray}
Extremizing this surface we find the saddle point equations
\begin{eqnarray}
\fl r,R:\qquad m = \int Dz \tanh(\sqrt{R} z + r) \qquad q = \int Dz
\tanh^2(\sqrt{R} z + r)\label{eq:saddle_mq}\\ 
\fl \hat{M},\hat{E}: \quad M = m \quad E = \frac{\alpha}{2}
\left\{\frac{\beta_s[1 + \hat{E}(1-q)^2] - 2m\hat{F}\beta_s(1-q) +
  \hat{F}^2(1-q)^2(1+\beta_s)}{\beta_s[1 + \hat{E}(1-q)]^2} \right\}\nonumber\\
\fl \hat{F} :\qquad F = \alpha \frac{\beta_s m - (1+\beta_s)
  \hat{F}(1-q)}{\beta_s [1  + \hat{E}(1-q)]} \label{eq:saddle_F}\\
\fl m,q:\qquad r + \hat{M} = \frac{-\alpha \hat{F}}{1 + \hat{E}(1-q)}
\qquad R = \frac{\alpha [\hat{F}^2(1 + \beta_s^{-1}) +
    \hat{E}(\hat{E}q + 2m\hat{F})]}{[1 + \hat{E}(1-q)]^2}\label{eq:saddle_Rr}
\end{eqnarray}
Equilibrium corresponds to $\hat{M} = 0$ and $\hat{E} = -\hat{F} =
\beta$ since in this case the magnetisation is unconstrained while the
variable conjugate to the energy is, of course, the temperature. It is a
useful check to see that in this case the saddle point equations for
$r,R, m$ and $q$ resort to their standard equilibrium expressions (as
then our maximum entropy measure is just the equilibrium measure at
temperature $\beta$). 

\subsection{Reduction of the saddle point equations}
We have to express the values of the unknown conjugate order
parameters given the values of the true order parameters. By
eliminating $\hat{F}$ from the saddle point equations for $F$ and $E$
we obtain a quadratic equation in $\hat{E}$ for which the physical
solution (utilising the fact that in equilibrium we require $\hat{E} =
\beta$) is given by 
\begin{eqnarray}
\fl \hat{E}(q) = \frac{1}{2d(1-q)^2} \bigg\{-[2d(1-q) + \beta_s(1+
  \beta_s)(1-q)^2] \label{eq:hatEofq} \\
\hspace*{-10mm}- \sqrt{[2d(1-q) + \beta_s(1+  \beta_s)(1-q)^2]^2 -
  4d(1-q)^2[d + \beta_s(1+\beta_s) - M^2 \beta_s^2]}\bigg\}\nonumber\\
d = \frac{F^2 \beta_s^2}{\alpha^2} - \frac{2 \beta_s (1 + \beta_s)
  E}{\alpha}\nonumber
\end{eqnarray}
Insertion of (\ref{eq:hatEofq}) into (\ref{eq:saddle_F}) gives us
$\hat{F}(q)$ which in turn implies $R(q)$. To obtain $r(q)$ we have to
use the implicit equation for $r$
\begin{equation}
M = \int Dz \tanh(\sqrt{R(q)} z + r)
\end{equation}
We have resolved all free parameters into bare functions of $q$
and hence have the relatively straightforward one dimensional problem
\begin{eqnarray}
q = \int Dz \tanh^2(\sqrt{R(q)}z + r(q))
\end{eqnarray}
which we solve numerically.

\section{Derivation of the force terms in the order parameter flow}

We are now in a position to calculate $D(h)$ and $D(f,h)$. We
abuse notation 
slightly by using $\hat{M},\hat{E},\hat{F},q$ when we mean their values
taken in the saddle point given the values of $M, E$ and $F$ rather than seeing them as
variables. With that taken into account, we may write
\begin{eqnarray}
D(h) = \lim_{n \to 0} \int \frac{\rmd \hat{h}}{2\pi}\rme^{\rmi h
  \hat{h}} \sum_{\bsigma} \sigma_0 \rme^{- \hat{M}
  \sum_{\alpha > 0} \sigma^0 \sigma^\alpha + \frac{\alpha}{2}\Xi(\bsigma)}\\
D(f,h) = \lim_{n \to 0} \int \frac{\rmd \hat{h}\rmd\hat{f}}{ (2\pi)^2}
\rme^{\rmi h \hat{h} + \rmi f \hat{f}} \sum_{\bsigma} \rme^{- \hat{M}
\sum_{\alpha > 0} \sigma^0 \sigma^\alpha + \frac{\alpha}{2}\Lambda(\bsigma)}
\end{eqnarray}
\begin{eqnarray}
\fl \Xi(\bsigma) =\frac{\int \rmd r \rmd \mathbf{v}
G(r, \mathbf{v}) \left[\beta_s(r - v^0) \sigma^0 -
 \sum_\alpha\sigma^\alpha ( \hat{F} r + \hat{E} v_\alpha)
-  \rmi \hat{h} (r-v^1)
  \right]^2 }{\int \rmd r \rmd \mathbf{v}  G(r, \mathbf{v})} \\
\fl \Lambda(\bsigma) = \frac{\int \rmd r \rmd \mathbf{v}
G(r, \mathbf{v}) \left[\beta_s(r - v^0) \sigma^0 -
 \sum_\alpha\sigma^\alpha ( \hat{F} r + \hat{E} v_\alpha)
-  \rmi \hat{h}v^1 - \rmi \hat{f}r
  \right]^2 }{\int \rmd r \rmd \mathbf{v}  G(r, \mathbf{v})} \label{eq:Lambda_def}\\
 G(r, \mathbf{v}) =  \rme^{-\frac12 \mathbf{v} \mathbf{q}^{-1}
  \mathbf{v} -\frac{\beta_s}{2} \left(r - 
  v^{0}\right)^2- \frac{\hat{E}}{2} \sum_\alpha v_\alpha^2 - \hat{F}
  \sum_\alpha v_\alpha r } \label{eq:G_measure}
\end{eqnarray}

\subsection{Evolution of the magnetisation}
\label{sec:mag_evol}
We concentrate first on $\Xi$. It is convenient to introduce the
shorthand for averages over the measure (\ref{eq:G_measure}), 
\begin{eqnarray}
\bra \ldots \ket_{G} = \frac{\int \rmd r \rmd \mathbf{v} G(r,
  \mathbf{v}) \ldots}{\int \rmd r \rmd \mathbf{v} G(r,
  \mathbf{v})}
\end{eqnarray}
with which we define the shorthand:
\begin{eqnarray}
\fl g_{10} = \alpha \beta_s \bra (r - v^0)(r - v^1) \ket_G \quad&
g_{0\alpha} = \alpha \beta_s\bra (r - v^0)(\hat{F} r + \hat{E} v^\alpha) \ket_G + \hat{M}\label{eq:factor1}\\
\fl g_{11} = \alpha \bra (r- v^1)^2\ket_G \quad&
g_{\alpha \beta} = \alpha \bra (\hat{F} r + \hat{E} v^\alpha)(\hat{F} r +
\hat{E} v^\beta) \ket_G\\
\fl g_{1\alpha} = \alpha \bra (r - v^1)(\hat{F} r + \hat{E} v^\alpha)
\ket_G \quad \forall \alpha > 1\quad&
g_{111} = \alpha \bra (r-v^1)(\hat{F}r + \hat{E} v^1) \ket_G\label{eq:factor2}
\end{eqnarray}
We calculate these factors in \ref{sec:app_Mg}, for now we
merely note that $g_{\alpha \beta} = R$ while $-g_{0\alpha}=r$.
It is possible to ignore constants such as $\beta_s^2\bra (r-v_0)^2 \ket_G$ since
they may be dealt with via overall normalization of the measure
$D(h)$. Then
\begin{eqnarray}
\fl D(h) \sim \lim_{n \to 0} \int \frac{\rmd \hat{h}}{2\pi}
  \rme^{-\frac{\hat{h}^2  g_{11}}{2} + \rmi
  h \hat{h}} \sum_{\bsigma} \sigma_0 \rme^{\rmi \hat{h}
[ \sigma^1 g_{111} + \sum_{\alpha>1} \sigma^\alpha g_{1\alpha} 
  - \sigma_0 g_{10} ]}
\int Dx \rme^{x \sqrt{R} \sum_\alpha
  \sigma^\alpha + \sigma_0 \sum_\alpha  \sigma^\alpha r }\nonumber
\end{eqnarray}
Since we only require $D(h)$ in a term of the form $\int \rmd h D(h)
\tanh(\beta h)$ it is possible to make the gauge transformation $x, h,
\hat{h} \to \sigma_0x, \sigma_0h,\sigma_0\hat{h}$ to remove the
dependence on $\sigma_0$. We then perform the trace over the other
spins to obtain
\begin{eqnarray}
\fl D(h) \sim \lim_{n \to 0} \int \frac{\rmd \hat{h}}{2\pi}
  \rme^{-\frac{\hat{h}^2 g_{11}}{2} + \rmi
  h \hat{h}} \int Dx \rme^{-\rmi \hat{h}
 g_{10}} \cosh[\rmi\hat{h} g_{111} +
  x\sqrt{R} +r]
\cosh^{n-1}[\rmi\hat{h} g_{1\alpha} + x\sqrt{R} +  r]\nonumber
\end{eqnarray}
In order to move our integration contour so that we can perform these
integrals we expand the first cosh function,
\begin{eqnarray}
\fl \cosh[\rmi \hat{h} g_{111} + x\sqrt{R} + r]\label{eq:sinh_fns} \\ 
= \cosh( \rmi \hat{h}\Delta)\cosh[\rmi\hat{h}
  g_{1\alpha} + x\sqrt{R} +r] +
 \sinh( \rmi \hat{h}\Delta)\sinh[\rmi\hat{h}
  g_{1\alpha} + x\sqrt{R} +r]\nonumber
\end{eqnarray}
where $\Delta =  g_{111} - g_{1\alpha} $. The integral over the
contribution to $D(h)$
containing only cosh functions is relatively straightforward and gives
\begin{eqnarray}
\fl \frac{1}{2\sqrt{2\pi
g_{11}}}\Bigg\{\exp\left[-\frac{(h - g_{10} + \Delta)^2}{2 g_{11}} \right] + 
\exp\left[-\frac{(h - g_{10} - \Delta)^2}{2 g_{11}} \right] \Bigg\}
\end{eqnarray}
For the term containing the sinh functions from (\ref{eq:sinh_fns}), we
shift the integration variable $x \to x - \rmi
\hat{h}\frac{g_{1\alpha}}{\sqrt{R}}$ and take the limit $n \to 0$
to obtain,
\begin{eqnarray}
\fl \int \frac{\rmd \hat{h}}{2\pi} \rme^{-\frac{ \hat{h}^2
 }{2}(g_{11} - \frac{g_{1\alpha}^2}{R})  + \rmi h
 \hat{h} -\rmi \hat{h} g_{10}} \sinh(\rmi \hat{h}\Delta)  \int
 Dx \rme^{\rmi x\hat{h} \frac{g_{1\alpha}}{\sqrt{R}}} \tanh(x\sqrt{R} +r)\nonumber
\end{eqnarray}
We proceed by writing $\sinh(\rmi \hat{h}\Delta) = \frac12(\rme^{\rmi
  \hat{h}\Delta} - \rme^{-\rmi\hat{h}\Delta})$ and we first treat the
  term with containing $\rme^{\rmi \hat{h} \Delta}$, integrating over $\hat{h}$:
\begin{eqnarray}
\fl \int \frac{\rmd \hat{h}}{4\pi} \rme^{-\frac{ \hat{h}^2
}{2}(g_{11} - \frac{g_{1\alpha}^2}{R}) +  \hat{h} \rmi (h - g_{10} + \Delta) } 
\int Dx \rme^{\rmi x\hat{h} \frac{g_{1\alpha}}{ \sqrt{R}}}
\tanh(x\sqrt{R} +r)\\
\fl = \frac{1}{2\sqrt{2\pi \alpha(g_{11} - \frac{g_{1\alpha}^2}{R}) }}
\int Dx \exp\left[-\frac{1}{2 (g_{11} -
    \frac{g_{1\alpha}^2}{g_{\alpha \beta}}) } (h  - g_{10} + \Delta + x \frac{g_{1\alpha}}{
 \sqrt{R}})^2\right] \tanh(x\sqrt{R} + r)\nonumber
\end{eqnarray}
By a careful change of variables this reduces to 
\begin{eqnarray}
  \frac{1}{\sqrt{2\pi g_{11}}} \rme^{-\frac{ h^2}{2
    g_{11}}}
\int Dx \tanh\left[x\sqrt{\frac{Rg_{11}   - g_{1\alpha}^2}{g_{11}}} -
   \frac{g_{1\alpha}}{g_{11}}h + r \right]
\end{eqnarray}
The case with $\rme^{-\rmi\Delta\hat{h}}$ follows identically, so
finally we find that
\begin{eqnarray}
\fl D(h) = \frac{\rme^{-\frac{1}{2
    g_{11}} (h - g_{10} + \Delta)}}{\sqrt{2\pi g_{11}}} \left\{1 + \int Dx
    \tanh\left[x\sqrt{\frac{ (Rg_{11}
   - g_{1\alpha}^2)}{g_{11}}} - \frac{g_{1\alpha}}{g_{11}}(h - 
   g_{10} + \Delta) + r \right]\right\}\nonumber\\
\fl + \frac{\rme^{-\frac{1}{2
    g_{11}} (h - g_{10} - \Delta)}}{\sqrt{2\pi g_{11}}} \left\{1 - \int Dx
    \tanh\left[x\sqrt{\frac{ (Rg_{11}
   - g_{1\alpha}^2)}{g_{11}}} - \frac{g_{1\alpha}}{g_{11}}(h - 
   g_{10} - \Delta) + r \right]\right\}
\end{eqnarray}
We wish to consider the term $\int \rmd h D(h) \tanh(\beta h)$ which
is required to calculate the force term in the differential equation (\ref{eq:M_ode}). This can
be most easily effected by first making the transformation $h \to h +
g_{10} \mp \Delta$ as required, followed by $h \to \sqrt{g_{11}} h$. We
then have
\begin{eqnarray}
\fl \int \rmd h D(h) \tanh(\beta h)\\
\fl= \frac12\int Dh Dx \left\{1 +  \tanh\left[x\sqrt{\frac{Rg_{11}
   - g_{1\alpha}^2}{g_{11}}} - \frac{g_{1\alpha}h}{\sqrt{g_{11}}} + r
    \right]\right\} \tanh[\beta(\sqrt{g_{11}} h + g_{10} - \Delta)] 
    \nonumber\\ 
\fl+ \frac12\int Dh Dx \left\{1 -
    \tanh\left[x\sqrt{\frac{Rg_{11}
   - g_{1\alpha}^2}{g_{11}}} - \frac{g_{1\alpha}h}{\sqrt{g_{11}}} + r
    \right]\right\} \tanh[\beta(\sqrt{g_{11}} h + g_{10} + \Delta)] 
    \nonumber\\ 
\end{eqnarray}
We then rotate the Gaussian integration variables which leads to our final result
\begin{eqnarray}
\fl \int \rmd h D(h) \tanh(\beta h)\\
\fl= \frac12\int Du Dv \left\{1 +  \tanh\left[\sqrt{R} u + r
    \right]\right\} \tanh[\beta(\sqrt{\frac{Rg_{11} - g_{1\alpha}^2}{R}}v -
\frac{g_{1\alpha}}{\sqrt{R}} u  + g_{10} - \Delta)] +
    \nonumber\\ 
\fl\frac12 \int Dh Dx \left\{1 -
    \tanh\left[\sqrt{R} u + r
    \right]\right\} \tanh[\beta( \sqrt{\frac{Rg_{11} - g_{1\alpha}^2}{R}}v -
\frac{g_{1\alpha}}{\sqrt{R}} u  + g_{10} + \Delta)] 
    \nonumber 
\end{eqnarray}

\subsection{Fixed points for the magnetisation in equilibrium}
As a useful test of our analysis thus far, we show that in
equilibrium, the equilibrium value of the magnetisation is a fixed
point of our dynamics. In equilibrium we have $\hat{M} = 0$ and
$\hat{E} = -\hat{F} = \beta$ as argued above. So we can read off the
equilibrium values of our factors $g_{\cdots}$ from 
\ref{sec:app_Mg}. In equilibrium $-\beta g_{1\alpha} =
R$ and $\beta g_{10} = r$. Further, $\beta g_{1\alpha} = -R$ while
$\beta g_{11} = - g_{111}$, whence 
\begin{eqnarray}
\beta \Delta = \beta (g_{111} - g_{1\alpha})  =R  - \beta^2 g_{11}
\end{eqnarray} and so
\begin{eqnarray}
\int \rmd h D(h) \tanh(\beta h) = \int Du \tanh[\sqrt{R} u + r]
\end{eqnarray}
according to the identity 
\begin{eqnarray}
\tanh(u) = &\frac12[1-\tanh(u)]\int Dy \tanh(u + yz - z^2)\nonumber\\
&+ \frac12[1 + \tanh(u)] \int Dy \tanh(u + yz + z^2)
\end{eqnarray}
of \cite{Coolen94a}. Since our saddle point equations
(\ref{eq:saddle_mq}-\ref{eq:saddle_Rr}) for
$m,q,R,r$ in equilibrium are equivalent to the equilibrium saddle point equations we see that in
equilibrium, one fixed point of our differential equation for the
magnetisation (\ref{eq:M_ode}) is given by the equilibrium magnetisation.

\subsection{Evolution of the energetic terms $E$ and $F$}

We now turn to $D(f,h)$, for which we need the evaluation of
$\Lambda(\bsigma)$ as defined in (\ref{eq:Lambda_def}). 
On top of our previous definitions we now introduce
\begin{eqnarray}
g_{rr} = \alpha \bra r^2 \ket_G\qquad &
g_{vv} = \alpha \bra v_1^2 \ket_G\label{eq:factor3}\\
g_{rv} = \alpha \bra r v_1 \ket_G\qquad&
g_{ra} = \alpha \bra r(\hat{F}r + \hat{E} v_\alpha) \ket_{G}\\
g_{va} = \alpha \bra v_1 (\hat{F}r + \hat{E} v_\alpha) \ket_{G}\quad
\forall \alpha > 1\qquad&
g_{v1} = \alpha \bra v_1 (\hat{F}r + \hat{E} v_1) \ket_{G}\\
g_{0r} = \alpha \beta_s\bra r (r - v_0) \ket_G\qquad&
g_{0v} = \alpha \beta_s \bra v_1 (r - v_0) \ket_G\label{eq:factor4}
\end{eqnarray}
which are also calculated in appendix \ref{sec:app_Mg}. Using these
$g$ factors we may write
\begin{eqnarray}
\fl D(f,h) \sim \lim_{n \to 0} \int \frac{\rmd \hat{h} \rmd
  \hat{f}}{(2\pi)^2 }
  \rme^{-\frac{\hat{h}^2  g_{vv}}{2} - \frac{\hat{f}^2 
  g_{rr}}{2} - \hat{f} \hat{h}  g_{rv}+ \rmi
  h \hat{h} + \rmi f \hat{f}} \sum_{\bsigma}  \rme^{\rmi \hat{h}
[\sigma^1 g_{v1}+ \sum_{\alpha>1} \sigma^\alpha
  g_{v\alpha} -  \sigma_0 g_{0v} ]}\nonumber\\
\int Dx \rme^{x \sqrt{R} \sum_\alpha
  \sigma^\alpha + r\sigma_0 \sum_\alpha  \sigma^\alpha} \rme^{-\rmi
  \hat{f} [ \sigma_0 g_{r0} -
 \sum_\alpha \sigma_\alpha g_{r\alpha}]}  
\end{eqnarray}
which we may rewrite as
\begin{eqnarray}
\fl D(f,h) \sim \lim_{n \to 0} \int \frac{\rmd \hat{h} \rmd
  \hat{f}}{(2\pi)^2 }
  \rme^{-\frac{\hat{h}^2 g_{vv}}{2} - \frac{\hat{f}^2 
  g_{rr}}{2} - \hat{f} \hat{h}  g_{rv}+ \rmi
  h \hat{h} + \rmi f \hat{f}} \int Dx \sum_{\sigma^0} \rme^{-\rmi
 \sigma_0 (\hat{h} g_{0v} 
 + \hat{f}g_{0r})}\label{eq:dfh_1}\\
   \cosh[\rmi\hat{h} g_{v1} + \rmi \hat{f} g_{r\alpha} + 
  x\sqrt{R} + \sigma^0 r]
\cosh^{n-1}[\rmi\hat{h} g_{v\alpha} +  \rmi \hat{f} g_{r\alpha}
  + x\sqrt{R} + \sigma^0 r]\nonumber 
\end{eqnarray}
Following a similar procedure to that used in $\S$ \ref{sec:mag_evol}
we write
\begin{eqnarray}
\fl \cosh[\rmi\hat{h} g_{v1} + \rmi \hat{f} g_{r\alpha} + 
  x\sqrt{R} + \sigma^0 r] &= &\cosh[-\rmi \hat{h}\Delta] \cosh[\rmi
  \hat{h} g_{v\alpha} +  \rmi \hat{f} g_{r\alpha} 
  + x\sqrt{R} + \sigma^0 r]\label{eq:cosh_dfh}\\ &+& \sinh[-\rmi
  \hat{h}\Delta] \sinh[\rmi \hat{h} g_{v\alpha} +  \rmi \hat{f}
  g_{r\alpha} 
  + x\sqrt{R} + \sigma^0 r]\nonumber
\end{eqnarray}
since $g_{v1} - g_{v\alpha} = -\Delta$. It is convenient to use the
change of variable $h \to h + g_{0v} \sigma_0 \pm \alpha \Delta$ and $f \to f +
g_{0r} \sigma_0$. The terms in (\ref{eq:dfh_1}) containing the cosh
terms from (\ref{eq:cosh_dfh}) can be treated in a similar manner to 
 $\S$ \ref{sec:mag_evol} and contribute
\begin{eqnarray}
\fl \frac{\sum_{\sigma_0}}{4\pi
  \sqrt{g_{rr}g_{vv} -
  g_{rv}^2}}\rme^{-\frac{g_{rr}h^2}{2(g_{rr} g_{vv} - 
  g_{rv}^2)} -\frac{g_{vv} f^2}{2(g_{rr} g_{vv} -
  g_{rv}^2)} + \frac{g_{rv} fh}{(g_{rr} g_{vv} -
  g_{rv}^2)}}
\end{eqnarray}
Now we tackle the more tricky sinh terms from (\ref{eq:dfh_1}) which
requires the shift of the $x$ integral in the complex plane but then gives 
\begin{eqnarray}
\fl \sum_{\sigma_0} \int \frac{\rmd \hat{h} \rmd
  \hat{f}}{2(2\pi)^2 }
  \rme^{-\frac{\hat{h}^2  g_{vv}}{2} - \frac{\hat{f}^2 
  g_{rr}}{2} - \hat{f} \hat{h} g_{rv}+ \rmi
  h \hat{h} + \rmi f \hat{f}} \int Dx \tanh[x\sqrt{R}  + \rmi (\hat{h}
  g_{v\alpha} + \hat{f} g_{r\alpha})+r\sigma_0]\nonumber \\
\fl = \sum_{\sigma_0} \int \frac{\rmd \hat{h} \rmd
  \hat{f}}{2(2\pi)^2 }
  \rme^{-\frac{\hat{h}^2 }{2}(g_{vv} -
  \frac{g_{v\alpha}^2}{R})  - \frac{\hat{f}^2 
}{2}(  g_{rr} - \frac{g_{r\alpha}^2}{R}) - \hat{f}
  \hat{h}( g_{rv} - \frac{g_{r\alpha} g_{v\alpha}}{R}) + \rmi
 \hat{h}(h +  \frac{xg_{v\alpha}}{\sqrt{R}})+
  \rmi \hat{f} ( f + \frac{x g_{r\alpha}}{\sqrt{R}})}\\
\times \int Dx  \tanh[x\sqrt{R} + r\sigma_0]\nonumber
\end{eqnarray}
Proceeding with the integrals over $\hat{h}$, $\hat{f}$ and after some
rearrangement we find,
\begin{eqnarray}
\fl \sum_{\sigma_0} \frac{1}{2\pi
  \sqrt{g_{rr}g_{vv} -
  g_{rv}^2}}\rme^{-\frac{g_{rr}h^2}{2(g_{rr} g_{vv} - 
  g_{rv}^2)} -\frac{g_{vv} f^2}{2(g_{rr} g_{vv} -
  g_{rv}^2)} + \frac{g_{rv} fh}{(g_{rr} g_{vv} -
  g_{rv}^2)}}\nonumber\\
\fl \hspace*{10mm} \times \int Dx \tanh[x\sqrt{\frac{R}{C}} - \frac{(g_{rr}
  g_{v\alpha} - g_{r\alpha} g_{rv})}{(g_{rr} g_{vv} - g_{rv}^2)} h -
  \frac{(g_{vv} 
  g_{r\alpha} - g_{v\alpha} g_{rv})}{(g_{rr} g_{vv} - g_{rv}^2)} f +r\sigma_0]
\end{eqnarray}
Putting this all together we have
\begin{eqnarray}
\fl \int \rmd f \rmd h D(f,h) f \tanh[\beta(f - h)] = \sum_{\sigma_0
  \sigma_1} \int \frac{\rmd f \rmd h}{4\pi 
  \sqrt{g_{rr}g_{vv} -  g_{rv}^2}} \nonumber\\
\fl \rme^{-\frac{g_{rr}h^2}{2(g_{rr} g_{vv} - 
  g_{rv}^2)} -\frac{g_{vv} f^2}{2(g_{rr} g_{vv} -
  g_{rv}^2)} + \frac{g_{rv} fh}{(g_{rr} g_{vv} -
  g_{rv}^2)}}\nonumber\\
\fl \Bigg\{1 + \sigma_1  \int Dx \tanh[x\sqrt{\frac{R}{C}} - \frac{(g_{rr} 
  g_{v\alpha} - g_{r\alpha} g_{rv})}{(g_{rr} g_{vv} - g_{rv}^2)} h -
  \frac{(g_{vv}  g_{r\alpha} - g_{v\alpha} g_{rv})}{(g_{rr} g_{vv} -
  g_{rv}^2)} f + r\sigma_0] \Bigg\} \nonumber\\ 
\fl (f + \sigma_0 g_{0r}) \tanh[\beta(f + g_{0r}\sigma_0 - (h +
  g_{0v}\sigma_0 + \sigma_1 \Delta))]
\end{eqnarray}
Rescaling our Gaussian variables in order to write the measures as
standard Gaussian measures we finally have the force term required for
(\ref{eq:F_ode}) 
\begin{eqnarray}
\fl \int \rmd f \rmd h D(f,h) f \tanh[\beta(f - h)] \\
=\frac14 \sum_{\sigma_0
  \sigma_1} \int Dy Dz [1 + \sigma_1 \int Dx
  \tanh(A + B)] C\tanh[\beta(C-D)] \nonumber
\end{eqnarray}
with
\begin{eqnarray}
A = x\sqrt{ \frac{R(g_{rr} g_{vv} - g_{rv}^2) - g_{v\alpha}^2 g_{rr} -
  g_{r\alpha}^2 g_{vv} + 2 g_{r\alpha} g_{v\alpha} g_{rv}}{(g_{rr}
  g_{vv} - g_{rv}^2)}} \\
B = r \sigma_0 - \sqrt{g_{rr}^{-1}}\left[\frac{g_{rr}g_{v\alpha} -
  g_{r\alpha} g_{rv}}{\sqrt{g_{rr} g_{vv} - g_{rv}^2}} + g_{r\alpha} y
  \right] \\
C = \sqrt{g_{rr}} y + g_{0r} \sigma_0\\
D =  \sqrt{\frac{(g_{rr} g_{vv} - g_{rv}^2)}{g_{rr}}}z + 
\sqrt{\frac{g_{rv}^2}{g_{rr}}} y + g_{0v}\sigma_0 + \Delta \sigma_1
\end{eqnarray}
while with these definitions, the force term required for
(\ref{eq:E_ode}) is 
\begin{eqnarray}
\fl \int \rmd f \rmd h D(f,h) h \tanh[\beta(f - h)] \\
=\frac14 \sum_{\sigma_0
  \sigma_1} \int Dy Dz [1 + \sigma_1 \int Dx
  \tanh(A + B)] D\tanh[\beta(C-D)] \nonumber
\end{eqnarray}

\section{Order parameter flow and comparison with simulations}
We now have a closed set of equations describing deterministic order
parameter flow. They are
\begin{eqnarray}
\frac{\rmd}{\rmd t} M = -M + \int \rmd h D(h) \tanh(\beta h)\\
\frac{\rmd}{\rmd t} E = -2E + \alpha + \int \rmd f \rmd h D(f,h) h
\tanh[\beta (f-h)] \\
\frac{\rmd}{\rmd t} F = -F + \int \rmd f\rmd h  D(f,h) f\tanh[ \beta(f-h)]
\end{eqnarray}
where $D(h)$ and $D(f,h)$ at any instant depend on the triple
$(M,E,F)$ (as well as the statistics of the quenched disorder via
$\beta_s$ and $\alpha$). 

\begin{figure}[t]
\vspace*{-2mm} \hspace*{45mm} \setlength{\unitlength}{0.75mm}
\begin{picture}(200,85)
\put(10,15){\epsfysize=60\unitlength\epsfbox{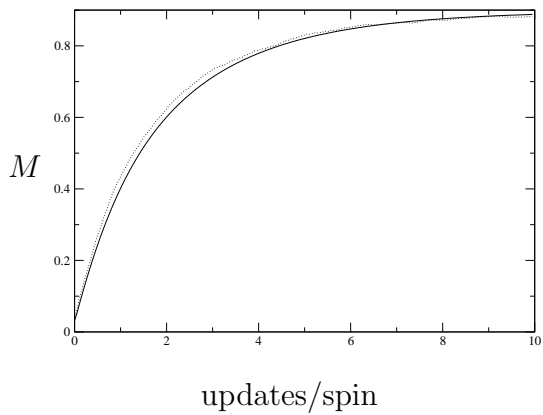}}
\put(52,6){\here{updates/spin}} \put(5,47){\here{$M$}}
\end{picture}
\vspace*{-8mm} \caption{We compare the results for the magnetisation
  $M$ of solving our order
  parameter equations numerically (solid line) with Monte Carlo
  simulations (dotted line) for MPM decoding at $\beta_s = \beta = 2$
  and $\alpha = 2$. The Monte Carlo simulations are performed with
  system size 2\,000 averaged over 50 samples (the standard deviation
  is less than $10^{-3}$).}
\label{fig:mag_evol}
\end{figure}

As an initial basic test of our theory, we compare the flow described
by the solution of our order parameter equations to Monte Carlo
simulations of the original Glauber dynamics in figure
\ref{fig:mag_evol}. The temperature is relatively high compared to
realistic values but we see that at least in this regime we have
excellent agreement between the theory and simulations, justifying our
assumptions and validating our method.

\subsection{Phase coexistence and basins of attractions}
It was noted by Tanaka \cite{Tanaka01, Tanaka02} that for certain
parameter regimes a spinodal would be encountered. This has drastic
implications for the decoding problem, since if there are metastable
states with a high bit error rate, local search algorithms will fail
to find the low bit error rate solution. To go beyond these arguments
a dynamic approach is required and we have now provided one. It is
possible to look at flow in parameter space and see explicitly the
basins of attraction for both the good and bad solutions. Spinodals
occur for both MAP decoding and MPM decoding. We focus on the latter
since the MPM decoding is optimal in terms of the bit error rate.
Since it is at Nishimori's temperature, for equilibrium states at least,
there are no complications due to replica symmetry breaking (we cannot
guarantee this for the dynamical saddle point at present, but we hope
to investigate this further in a later work). 

There are a variety of initial conditions with which we could start,
however two are of particular note. The first is the random initial
state (i.e. $\sigma_i(0) = \pm 1$ with probability $\frac12$). It is
straightforward to derive the initial states of our order parameters
as
\begin{eqnarray}
M_{RIS} = 0 \qquad F_{RIS} = 0 \qquad E_{RIS} = \frac{\alpha}{2}
\end{eqnarray}
A second important initial state is that given by the conventional
demodulator \cite{Verdu98}, with $\sigma_i(0) = \sgn(f_i)$. We derive
the values of our order parameters in appendix \ref{sec:app_CD}, which
are
\begin{eqnarray}
\hspace*{-12mm} M_{CD} = \mbox{Erf}\left[\sqrt{\frac{\alpha}{2(1 + \beta_s^{-1})}}\right]
  \qquad 
F_{CD} = \int Dz \left| \alpha + \sqrt{\alpha(1 + \beta_s^{-1})} z
  \right|\\
\hspace*{-12mm} E_{CD} = \frac{\alpha}{2} \left[1 + 2M_{CD} \chi + \chi^2(1 +
  \beta_s^{-2}) \right] \qquad \chi = \sqrt{\frac{2}{\pi\alpha(1 +
  \beta_s^{-1})}} \rme^{-\frac{\alpha}{2(1 + \beta_s^{-1})}}
\end{eqnarray}
where Erf$(x) = \frac{2}{\sqrt{\pi}} \int_0^x \rmd t \rme^{-t^2}$. 

\begin{figure}[t]
\vspace*{-2mm} \hspace*{45mm} \setlength{\unitlength}{0.75mm}
\begin{picture}(200,85)
\put(10,15){\epsfysize=60\unitlength\epsfbox{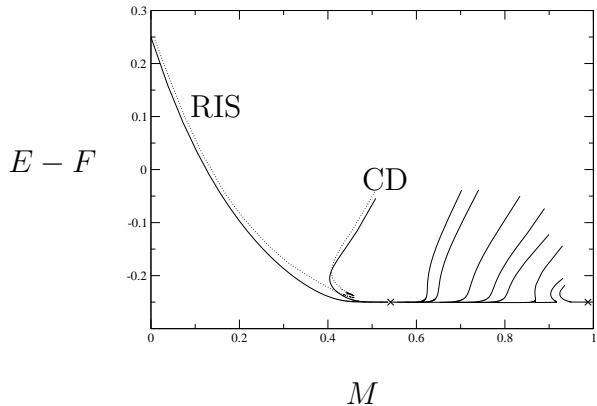}}
\put(52,6){\here{$M$}} \put(-3,47){\here{$E - F$}}
\put(26, 57){\here{RIS}} \put(56, 44){\here{CD}}
\end{picture}
\vspace*{-8mm} \caption{We examine the flow through phase space
  projected onto the energy (E-F) - overlap (M) plane. We both solve our
order parameter equations numerically (solid line) and compare
  against Monte Carlo simulations (dotted line) for the two important
  initial staes CD and RIS. We work with MPM decoding at $\beta_s = \beta = 17$
  and $\alpha = 0.5$. The crosses mark the two stable solutions to the
  equilibrium problem. The Monte Carlo simulations are performed with
  system size 2\,000 averaged over 50 samples. We ran both simulations
  and our theory for 1\,000 updates per spin. We see that for this
  parameter regime there is phase coexistence. The line labelled RIS
  starts from a random initial state, while the one labelled CD has
  the conventional demodulator as its initial state. Both flow into
  the poor attractor. We also show flow starting from some other (non-physical)
  initial states to get a better idea of the basins of attraction for
  the two phases. Note that all flow is from top to bottom as the
  dynamics lowers the energy of the system.}
\label{fig:emflow}
\end{figure}

In figure \ref{fig:emflow} we plot the flow of the dynamics in the
energy - overlap plane. We see that as expected the random initial
state (RIS) and conventional demodulator (CD) state both flow into the poor
attractor. The agreement between theory and simulations is quite
reasonable up to some low energy point where we believe replica
symmetry breaking (RSB) starts to play an important role in the
simulations (although our current replica symmetric theory is unable
to take it into account).
We also note that starting from the CD state the overlap
gets worse before improving to a
state better than the CD in the theoretical approach, while in the
simulations, it never recovers its initial overlap. Thus, starting from CD and running a local
search algorithm can significantly decrease
performance levels in a practical setting. We also see that the basin of attraction for the
good final state is significantly smaller than that for the poor final
state, the boundary is well over half way between the two states in
($E-F, M$) space. This also shows how distance in ($E-F, M$) space is
an unreliable guide to the direction of flow towards attractors.

\section{Conclusions}

CDMA is an important standard used in modern mobile
communications. Tools from statistical physics have provided and will
continue to provide useful ways of examining the detection
problem. Here we have developed and used dynamical replica theory to study the
dynamics of the detection problem for a prototypical local search
algorithm, namely the Gibbs sampler under Glauber dynamics. Although
this approach is only an analytic approximation, it provides a useful
counterpart to both density evolution and generating functional
analysis as a tool for examining dynamic rather than equilibrium
properties. As we have seen in comparison with Monte Carlo simulations
the approximation is a reasonably good one. We have also calculated
the basin of attraction for a particular set of parameters, in the
region where there is phase coexistence. As expected we have seen that
the practically available initial states, CD and RIS, both flow to the
poor solution. We have also seen that in this case the overlap
decreases initially from the CD state with our search algorithm and
that the basin of attraction for the good solution is relatively small
in energy-overlap space. One obvious extension of this work would be
to increase the order parameter set to improve the level of
approximation. Since the number of updates per spin required to visit
interesting regimes is of the order of $10^3$, this would have to be
done in a way that was compatible with practical numerical
solution. Another interesting problem is examining the role of replica
symmetry breaking on the dynamics. Finally, a project that we are
already working on is using the dynamical replica approach for
parallel update dynamics where we can compare its predictions to the
exact theory of generating functional analysis \cite{Mimura05}.

\appendix
\section{Calculation of factors defining $D(h)$ and $D(f,h)$}
\label{sec:app_Mg}
The various factors defined in (\ref{eq:factor1}-\ref{eq:factor2},
\ref{eq:factor3}-\ref{eq:factor4}) are all simple combinations of
\begin{eqnarray}
\bra r^2 \ket_G \qquad \bra r v^0 \ket_G \qquad \bra r v^\alpha \ket_G
\qquad \bra v^0 v^\alpha \ket_G \qquad
\bra v^\alpha v^\beta \ket_G \qquad \bra v^\alpha v^\alpha \ket_G
\end{eqnarray}
which are moments of the measure $G(r,\mathbf{v})$ (\ref{eq:G_measure})
The calculation is not dissimilar to that carried out in the
equilibrium calcultion; we have to just be careful with the algebra
over various Gaussian integrals. We find,
\begin{eqnarray}
\bra v^\alpha v^\alpha \ket_{G} &= &\frac{1-q}{1 + \hat{E}(1-q)}  + \frac{q +
(1-q)^2\hat{F}^2 (1 + \beta_s^{-1}) - 2(1-q) \hat{F}m }{[1 +
    \hat{E}(1-q)]^2}\\
\bra r^2 \ket_G &=& (1 + \beta_s^{-1})\\
\bra r v^0 \ket_G &=& 1\\
\bra r v^\alpha \ket_G &=& \frac{m - (1-q) \hat{F}(1 + \beta_s^{-1})}{[1 +
\hat{E}(1-q)]}\\
\bra v^0 v^\alpha \ket_G &=& \frac{m - (1-q) \hat{F}}{[1 +
\hat{E}(1-q)]}\\
\bra v^\alpha v^\beta \ket_G &=& \frac{q +
(1-q)^2\hat{F}^2 (1 + \beta_s^{-1}) - 2(1-q) \hat{F}m }{[1 +
    \hat{E}(1-q)]^2}
\end{eqnarray}
Hence the various factors are simply
\begin{eqnarray}
g_{10}  = \alpha \frac{1 + (1-q)(\hat{E} + \hat{F})}{[1 +
    \hat{E}(1-q)]} \qquad g_{0\alpha} = r \qquad g_{\alpha \beta} = R
\end{eqnarray}
and
\begin{eqnarray}
\fl g_{11} = \alpha \frac{\beta_s^{-1} + 2 - 2m +\hat{E}(1-q)^2 + 2(1-q)(\hat{E} +
  \hat{F})(\beta_s^{-1} + 1 - m) + (\beta_s^{-1} + 1)(1-q)^2(\hat{E} +
  \hat{F})^2} {[1 +
    \hat{E}(1-q)]^2}\nonumber
\end{eqnarray}
We then have
\begin{eqnarray}
\fl g_{1\alpha}= \alpha \frac{(1 + \beta_s^{-1})\hat{F} + (\hat{E} - \hat{F}) m -
  \hat{E} q + (\hat{E} + \hat{F})(1-q)(\hat{E} m + \hat{F}(1 +
  \beta_s^{-1}))}{[1 + \hat{E}(1-q)]^2}
\end{eqnarray}
which is  $-R$ in equilibrium. Then
\begin{eqnarray}
\fl \Delta = \frac{-\alpha\hat{E}(1-q)}{1 + \hat{E}(1-q)}
\qquad g_{v\alpha} = \alpha \frac{\hat{F} m + \hat{E} q - (1-q)[m \hat{E} \hat{F} + \hat{F}^2(1
    + \beta_s^{-1})]}{[1 + \hat{E}(1-q)]^2}
\end{eqnarray}
and 
\begin{eqnarray}
\fl g_{r\alpha} = \alpha\frac{\hat{F}(1 + \beta_s^{-1}) + \hat{E} m}{1 + \hat{E}(1-q)}
\qquad 
g_{r0} = \alpha \beta_s (\bra r^2 \ket_G  -\bra r v_0 \ket_G) =
\alpha\\
g_{v0} =\alpha \beta_s (\bra r v_1 \ket_G - \bra v_1 v_\alpha \ket_G)
\end{eqnarray}
Some of these factors only appear in certain combinations, however, although we
have tried we have not made any significant simplification through further
algebraic manipulation.

\section{Calculation of our order parameters for the conventional
  demodulator}
\label{sec:app_CD}
In this appendix, to lighten notation slightly we assume that the
original message was $\bsigma$ while our estimator is
$\hat{\bsigma}$. The conventional demodulator sets $\hat{\sigma}_i =
\mbox{sgn}(f_i)$ with  
\begin{equation}
f_i = \alpha \sigma_i + \frac1N \sum_{t,j\neq i} \eta_i^t \eta_j^t
\sigma_j + \frac1N \sum_t \eta_i^t \nu^t \label{eq:fidef}
\end{equation}
Since $\nu^t \sim \mathcal{N}(0, N/\beta_s)$ and we assume that
$\{\eta_i^t,\sigma_i,\nu^t\}$ are all mutually independent random
  variables, we can use the central limit theorem to treat the second
  and third terms in (\ref{eq:fidef}). So the error probability for a
  single bit is the probability that a Gaussian $\mathcal{N}(0,1)$
  random variable is less than $-\sqrt{\alpha/(1 + \beta_s^{-1})}$ which gives
\begin{eqnarray}
M_{CD} = \mbox{Erf} \left[\sqrt{\frac{\alpha}{2(1 + \beta_s^{-1})}} \right]
\end{eqnarray}
We can also see that $F_{CD} = \frac1N \sum_i f_i \mbox{sgn}(f_i) =
\mathbb{E}|f_i|$ is given by
\begin{eqnarray}
F_{CD} = \int Dz \left|\alpha + \sqrt{\alpha(1 + \beta_s^{-1})} z\right|
\end{eqnarray}
Now, we can write $E_{CD}$ as
\begin{eqnarray}
E_{CD} = \frac12 \sum_t ( m_t^{CD})^2 \qquad m_t^{CD} = \frac1N
\sum_i \eta_i^t \hat{\sigma}_i
\end{eqnarray}
Now,
\begin{eqnarray}
 \eta_i^s \hat{\sigma}_i = \eta_i^s \mbox{sgn}[\alpha \sigma_i + \frac1N \sum_{t,j\neq i} \eta_i^t \eta_j^t
\sigma_j + \frac1N \sum_t \eta_i^t \nu^t]\\
 = \eta_i^t \mbox{sgn}[\alpha \sigma_i + \frac1N \sum_{t\neq s,j\neq i} \eta_i^t \eta_j^t
\sigma_j + \frac1N \sum_{t\neq s} \eta_i^t \nu^t]\\ 
+ \frac{1}{N}\sum_i
\left(\sum_j \eta_j^s \sigma_j + \nu^s \right) 2\delta\left(\alpha
\sigma_i + \frac1N \sum_{t\neq s,j\neq i} \eta_i^t \eta_j^t 
\sigma_j + \frac1N \sum_{t\neq s} \eta_i^t \nu^t \right) + \mathcal{O}(N^{-1})\nonumber
\end{eqnarray}
where we have thrown away irrelevant terms and expanded the sgn
function in a Taylor expansion (it is possible, although longer to
check we obtain the same result without using this trick). We define
\begin{eqnarray}
\gamma^t = \frac1N \sum_i \eta_i^t\mbox{sgn}[\alpha \sigma_i + \frac1N \sum_{t\neq s,j\neq i} \eta_i^t \eta_j^t
\sigma_j + \frac1N \sum_{t\neq s} \eta_i^t \nu^t]\\
\chi = \frac{1}{N}\sum_i 2\delta\left(\alpha
\sigma_i + \frac1N \sum_{t\neq s,j\neq i} \eta_i^t \eta_j^t 
\sigma_j + \frac1N \sum_{t\neq s} \eta_i^t \nu^t \right)\\
= \frac12 \sum_\sigma \int Dz 2 \delta[\alpha\sigma + \sqrt{\alpha (1
    + \beta_s^{-1})} z] = \sqrt{\frac{2}{\pi\alpha(1 +
  \beta_s^{-1})}} \rme^{-\frac{\alpha}{2(1 + \beta_s^{-1})}}\\
\kappa^t = \frac{1}{N} \sum_j \eta_j^t \sigma_j
\end{eqnarray}
So up to irrelevant constants, 
\begin{eqnarray}
m_t^{CD} = \gamma^t + \kappa^t \chi + \sqrt{N\beta_s^{-1}} \chi z_1
\end{eqnarray}
where above and in the following $z_1, z_2, z_3$ are
$\mathcal{N}(0,1)$ random variables. The slight complication is that
$\gamma^t$ and $\kappa^t$ are correlated since the latter is a trace over
$\eta_i^t \sigma_i$ while the former is a trace over essentially
$\eta_i^t \hat{\sigma}_i$ - thus $M_{CD}$ determines the degree of correlation
and we find
\begin{eqnarray}
E_{CD} = \frac{\alpha}{2} \left[1 + 2M_{CD} \chi + \chi^2(1 +
  \beta_s^{-2}) \right]
\end{eqnarray}

\section*{Acknowledgments}
The research of MO is partially supported by
Grant-in-Aid No. 14084212 and No. 16500093
from the Ministry of Education, Culture,
Sports, Science, and Technology, the Japanese Government.
JPLH would like to thank M. Yoshida and T. Uezu for helpful
discussions on the CDMA detection problem.

\end{document}